\documentclass[prb,reprint]{revtex4-1}
\usepackage{H1}
\usepackage{amsmath}
\usepackage[pdftex,colorlinks=true]{hyperref}
\usepackage{amssymb}
\usepackage{amsthm}
\usepackage{amsfonts}
\usepackage{bbm}
\usepackage{array}
\newcolumntype{L}[1]{>{\raggedright\let\newline\\\arraybackslash\hspace{0pt}}m{#1}}
\newcolumntype{C}[1]{>{\centering\let\newline\\\arraybackslash\hspace{0pt}}m{#1}}
\newcolumntype{R}[1]{>{\raggedleft\let\newline\\\arraybackslash\hspace{0pt}}m{#1}}

\newcolumntype{N}{@{}m{0pt}@{}}

\begin{document}
\title{ Phase Structure of 1d Interacting Floquet Systems II: Symmetry-Broken phases }
\author{C.W.~von~Keyserlingk}
\author{S.~L.~Sondhi}
\begin{abstract}
Recent work suggests that a sharp definition of `phase of matter' can be given for periodically driven `Floquet' quantum systems exhibiting many-body localization. In this work we propose a classification of the phases of interacting Floquet localized systems  with (completely) spontaneously broken symmetries -- we focus on the one dimensional case, but our results appear to generalize to higher dimensions. We find that the different Floquet phases correspond to elements of $Z(G)$, the centre of the symmetry group in question. In a previous paper we offered a companion classification of unbroken, i.e., paramagnetic phases.
\end{abstract}
\maketitle

\section{Introduction}\label{s:introduction}
In this, the second of a series of two papers, we continue our investigation of Floquet drives which are many body localized (having a near complete set of bulk integrals of motion) and which also exhibit eigenstate order thus sharply defining phases for driven quantum systems. In our first paper  \cite{vonKeyserlingkSondhi16a}, henceforth I, we discussed Floquet drives with unbroken quantum order, which includes drives with  symmetry protected topological (SPT) order  (see Refs.~\onlinecite{Else16,Potter16,Harper16} for closely related work). In the present work we discuss Floquet drives with spontaneous symmetry broken (SSB) order.

 Here we briefly summarize the logic that leads to I and the present work --  we direct readers to the Introduction to I for a more complete account of the setting of this work, as well as indirectly related references. We consider many body Floquet localized systems whose eigenstates are non-thermal. In particular -- in part to avoid the runaway heating effect discussed in Refs.~\onlinecite{Lazarides14PRE,Abanin14,Ponte15} -- we assume that such localized Floquet systems are characterized by a set of commuting local integrals of the motion (which we call `l-bits') of the same form as those characterizing undriven many body localized systems\cite{Serbyn13a,Serbyn13cons,Huse14}. Thus the task of classifying possible Floquet phases reduces to classifying commuting stabilizer\footnote{Commuting stabilizer unitaries are unitaries with an extensive number of local conserved quantities. See  \secref{s:IsingPhaseDiagram} for examples.} unitaries much as the task of classifying possible MBL phases reduces to classifying commuting stabilizer Hamiltonians\cite{Potter15}.

In this paper we carry out this task for symmetry broken phases. Further we consider only finite symmetry groups $G$ -- for continuous groups, spontaneously broken states have Goldstone bosons which cannot be localized\cite{Gurarie03}. For simplicity we further restrict to finite on-site unitary symmetry groups $G$, and assume that $G$ has been completely spontaneously broken. In this task we again begin with the results obtained in Ref.~\onlinecite{Khemani15} which analyzed one dimensional driven spin chains with Ising/$Z_2$ symmetry and showed that they exhibited two symmetry broken phases, one of which has no analogue in the undriven setting. While we will improve upon the analysis in Ref.~\onlinecite{Khemani15} in understanding the structure and generality of the results obtained there and then generalize to arbitrary finite groups, we will rely on the computational evidence assembled there to argue that our classification describes stable Floquet phases of matter.

We can summarize our main result: We find that for MBL Floquet drives with a completely (spontaneously) broken on-site finite unitary symmetry group $G$,  the distinct Floquet drives are in correspondence with the elements $Z(G)$ of the center of the group. Unlike in I where for unbroken (i.e., SPT) phases we considered systems with edges, our classification here is done for a bulk system and the presence or absence of edges is not important. We show that the full period unitary $U(T)$ has a specific structure that reflects the center of the group. We also show that the evolution of the correlations inside the period have a characteristic form that derives from this structure on the lines discussed in Ref.~\onlinecite{Khemani15} for the Ising case.

\begin{table}
\begin{tabular}{|c|c|}
\hline 
$G$ & Classification $(Z(G))$\tabularnewline
\hline 
$\mathbb{Z}_{n}$ & $\mathbb{Z}_{n}$\tabularnewline
\hline 
$Q_{8}$ & $\mathbb{Z}_{2}$\tabularnewline
\hline 
$D_{n>2}$ & $\mathbb{Z}_{\frac{3+(-1)^{n}}{2}}$\tabularnewline
\hline 
$S_{n>2}$ & $\mathbb{Z}_{1}$\tabularnewline
\hline 
\end{tabular}
\caption{This table gives examples of our proposed $Z(G)$ classification scheme
for MBL Floquet drives in 1d with finite on-site symmetry group $G$,
and fully SSB eigenstate order. $\mathbb{Z}_1$ denotes the trivial group with one element. Only certain many-body localizable\cite{Potter15}
eigenstate orders are expected to persist in the Floquet setting\cite{Khemani15}.
For this reason we restrict attention to SSB orders with finite $G$
(see discussion in \secref{s:introduction}).}\label{tab:examples}
\end{table}

\tabref{tab:examples} gives examples of predictions arising from our framework. Groups $G$ with trivial centre -- e.g., the symmetric group $S_n$ on $n\geq 3$ elements, or the odd dihedral groups $D_{2 m+1}$ for $m\geq 1$ --  have just one Floquet phase with completely SSB order. The opposite extreme is abelian $G$, for which there are $|G|$ different Floquet fully SSB phases. On the other hand, the even dihedral groups $G=D_{2 n}$ have a  $\mathbb{Z}_2$ classification. 

This work is set out as follows. In \secref{s:IsingPhaseDiagram}, we investigate Floquet drives with Ising symmetry broken order, verifying that there are two qualitatively different such drives. In \secref{s:towards} we bring together some of our observations in this special case, discuss the stability of the Ising Floquet phases, and generalize them to drives with $\mathbb{Z}_n$ completely SSB order, showing that there are $n$ qualitatively different drives. Then in \secref{s:nonab} we extend all of these observations to consider fully symmetry broken orders for general finite $G$, and discuss the stability of these phases in general. Here we find $|Z(G)|$ qualitatively distinct Floquet phases.  In \secref{s:nonabexamples} we give a general prescription for constructing drives which realize the predicted Floquet phases. Finally in \secref{s:algebraic} we reflect on the structure of the spectra of these Floquet phases and make a connection to recent work on `time crystals'\cite{Wilczek12}, before concluding in \secref{s:conclusion}.

\section{Motivating Example: Ising chains}\label{s:IsingPhaseDiagram}
\begin{figure}[h]
\includegraphics[width=0.95\columnwidth]{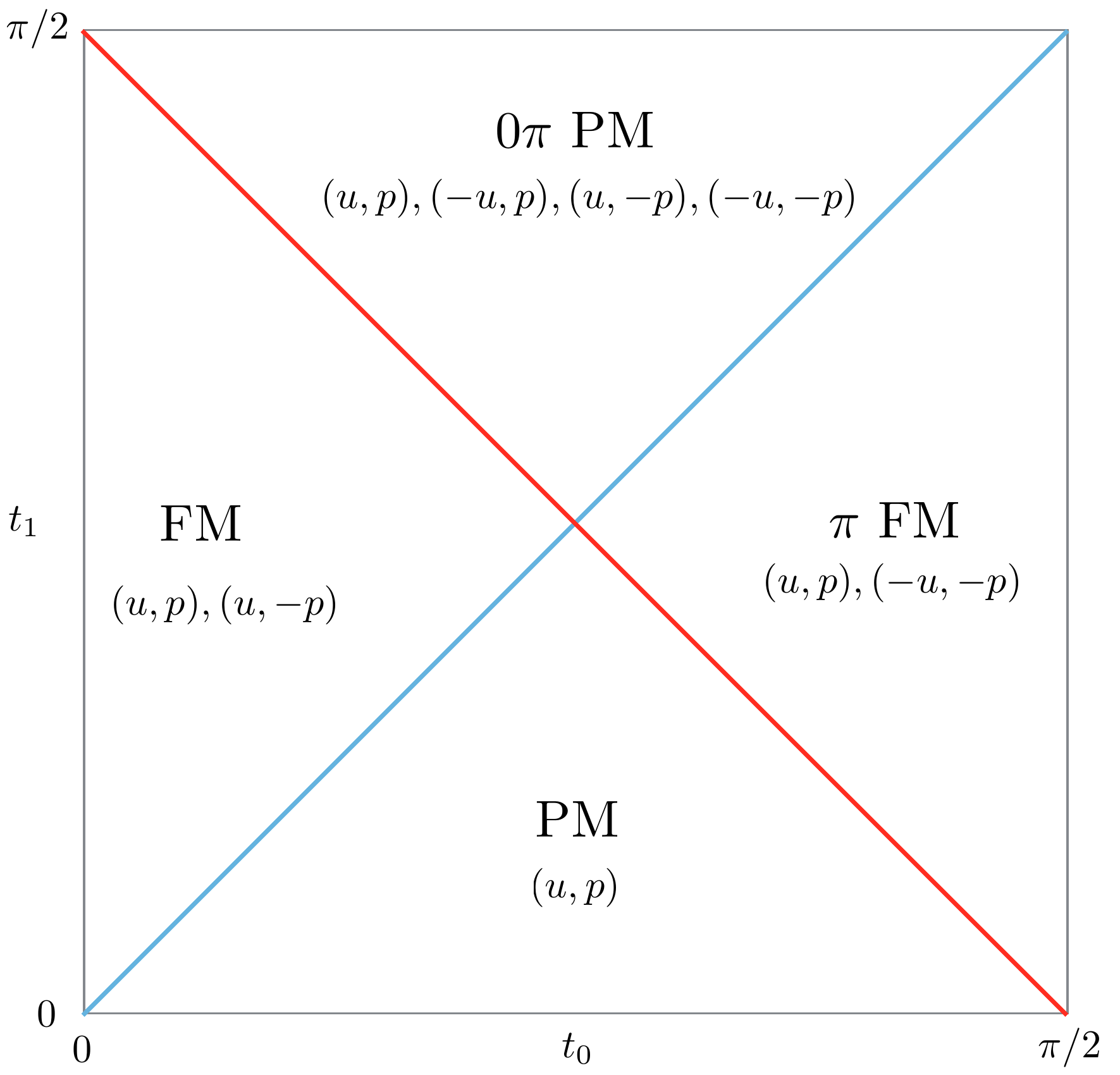}
\caption{(Color Online): This shows the phase diagram for the binary drive in \eqnref{eq:classDUf}. The red and blue line separate distinct Floquet phases. The lists involving $(u,p)$ summarize the protected multiplets in the spectrum for an open chain e.g., in the $0 \pi$ trivial phase, if there is a state with $U_f,P$ eigenvalues $(u,p)$  then there are guaranteed to be states at $(-u,p),(u,-p),(-u,-p)$ for the same $u$, up to exponentially small corrections in system size. } 
\label{classDPhasediagram}
\end{figure}

We begin with a discussion of our motivating case---that of spin chains with Ising symmetry. This was first discussed in Ref.~\onlinecite{Khemani15} and in more detail in I as the case of Class D fermionic chains. Our discussion will be mostly in the character of a review albeit with many details filled in.

\subsection{Solvable binary drives}

To this end we consider a set of binary drives which are exceptionally convenient and lead to the phase diagram \figref{classDPhasediagram} with two paramagnetic and two ferromagnetic phases\cite{Thakurathi13,Khemani15,vonKeyserlingkSondhi16a}. Indeed, they are spin versions of the Class D free fermion drives considered in I.
The drives are constructed from the Hamiltonians
\begin{align*}
H_{0}= & -\sum_s  h_{s}X_s \\
H_{1}= & -\sum_s J_{s}Z_s Z_{s+1}\punc{,}
\end{align*}
where $X,Z$ are Pauli-matrix operators and $h_{s},J_{s}$ are made random to obtain localization, but for the purposes of obtaining the phase boundaries will be taken to be (almost) spatially uniform. Both Hamiltonians commute with a $\mathbb{Z}_2$ global Ising symmetry operator $P=\prod_s X_s$. 
 
 $H_0$ is a paramagnetic fixed point Hamiltonian; its eigenstates exhibit zero correlation length in the Ising order parameter $Z_s$. It has a complete set of $N$ local conserved quantities (or l-bits) $\{X_s\}$ for a chain of length $N$, which themselves make up the terms in the Hamiltonian and are Ising symmetric. On the other hand, $H_1$ is a ferromagnetic fixed point Hamiltonian. Its eigenstates resemble classical configurations of the Ising order parameter $Z_s$, and indeed have perfect long range correlations in this order parameter. We say that $H_1$ has $\mathbb{Z}_2$ SSB eigenstate order because its eigenstates break the $\mathbb{Z}_2$ symmetry. $H_1$ is somewhat special in that its eigenstates can be chosen to be eigenstates of the $Z_s$. For a more general SSB Hamiltonian, the exact eigenstates of the system come in nearly degenerate doublets of feline/cat states, consisting (roughly) of  symmetric and antisymmetric combinations of Ising reversed order parameter configurations. Such pairs of cat states are labelled in part by l-bits of the form $B_s=Z_s Z_{s+1}$ which track the positions of domain walls. Only $N-1$ of these l-bits are independent for an $N$ site chain. The one additional integral of the motion needed to specify the eigenstates is the global Ising generator $P$ which commutes with $H_1$ and all the $B_s$ -- $P=\pm 1$ determine (roughly) whether or not the cat state is a symmetric or anti-symmetric combination of spin configurations. The operator $Z_r$ for any site $r$ can be used to toggle between these cat states, because it commutes with $H_1$ and anti-commutes with $P$. In summary, $B_s,P$ are a complete set of integrals of motion for Hamiltonians with SSB Ising order, although $P$ clearly cannot appear in any local Hamiltonian.

%so they are simultaneously diagonalizable $\mid E,p\rangle$. However, $Z_s$ (for any $s$) commutes with $H_1$ and anti-commutes with $V$, so $Z_{s}\mid E,v\rangle$ is a nonzero orthogonal state with $H_1=E,V=-v$. Thus the eigenstates of $H_1$ come in degenerate pairs with different Ising symmetry eigenvalues. 

Following Ref.~\onlinecite{Khemani15} (see also Ref.~\onlinecite{vonKeyserlingkMoessner16}) we now define binary Floquet drives using the reference Hamiltonians $H_{0},H_{1}$:
\[
U(t)=\begin{cases}
e^{-iH_{0}t} & 0\leq t<t_{0}\\
e^{-iH_{1}(t-t_{1})}e^{-iH_{1}t_{0}} & t_{0}\leq t<t_{0}+t_{1} \punc{.}
\end{cases}
\]
The final Floquet unitaries are of the form  
\begin{equation}\label{eq:classDUf}
U_f\equiv U(T)=e^{- i H_1 t_1 } e^{- i H_0 t_0 } \punc{.}
\end{equation}
  
If we choose spatially uniform couplings $h_s=J_s=1$ the phase transition lines are easily derived and lead to a phase diagram with four distinct phases as in \figref{classDPhasediagram}. In reality we will need disorder for a proper realization of the phases listed thereon but much can be learned by simply working on the boundaries of the diagram. For the same reasons as explained in I, it suffices to consider Floquet drives with $t_0,t_1\in [0,\pi/2]$ (see \figref{classDPhasediagram}). We will primarily be interested in those regions labelled FM, as our previous work covers the cases with paramagnetic bulk order, but we discuss the latter as well for completeness (\secref{ss:paramagnetic}).

\subsubsection{Ferromagnetic Phases}\label{ss:FMregions}
In the region labelled FM, all of the eigenstates have long range Ising symmetry broken order. A representative Floquet unitary is obtained by setting $t_0=0$ i.e., $U_f=e^{-i H_1 t_1}$. The eigenstate properties of this unitary are simply those of the local Hamiltonian $H_1$ which is the so-called Floquet Hamiltonian for this drive, i.e. the logarithm of $U_f$. Clearly, the l-bits which commute with $U_f$ are then the set $B_s$ and there is one global integral of the motion $P$. Note that $P$ does not appear in $U_f$. This unitary inherits the spectral pairing characteristic of the SSB broken order present in the eigenstates of $H_1$. That is to say, as $Z_r$ commutes with $H_1$ but anti-commutes with $P$, eigenstates at a given quasienergy come in $P=\pm 1$ pairs. The time
dependence of the order parameter in the Floquet eigenstates is also interesting---it returns to itself at the end of the period \cite{Khemani15}.

The $\pi \text{FM}$ phase is unique to the driven setting. To understand the nature of this phase, we work along the line $t_0 = \pi/2$ and $0<t_1 < \frac{\pi}{2}$ where 
$$
U_f = e^{-  i t_1 H_1} \prod_{s} X_s \propto e^{-  i t_1 H_1} P \ .
$$
This $U_f$ looks like the unitary associated with a FM drive (discussed above) multiplied by $P$.  Observe that $U_f$ is {\it not} the exponential of a local Hamiltonian although it is a local unitary generated by a local time dependent Hamiltonian. The complete set of integrals of the motion are again the $B_s$ and P but now $P$ {\it does} appear in $U_f$. The consequence of this last fact is that whereas in the FM case considered above, there was spectral pairing at any given quasienergy, there is now spectral pairing between states split by quasienergy $\pi$. That is to say, as $Z_s$ anti-commutes with both $U_f$  and $P$, eigenstates come in $(u,p),(-u,-p)$  pairs as opposed to $(u,p),(u,-p)$ pairs. 

The fact that $Z_l$ anti-commutes with $U_f$ in the $\pi$ FM case can be reinterpreted as a dynamical statement about the order parameter $Z_l$ -- namely, that it changes sign over the course of a Floquet cycle. This observation formed the basis for the spin-correlation based diagnostic reported in Ref.~\onlinecite{Khemani15}. Thinking just about our family of binary drives it would appear that what is at issue is a rotation of the order parameter about the $x$-axis and thus there may exist Floquet phases corresponding to rotation of the $Z_l$ order parameter for any multiple of $\pi$. However this is not the case for more general drives---in \appref{app:2pirot} we show that only the angle of rotation modulo $2\pi$ matters, which is consistent with our seeing only two distinct Floquet phases. Finally, we note we were able to distinguish the FM phases by looking just at their bulk spectra -- there was no need to examine their edge spectra. We will later argue in \secref{ss:nothingnewattheedge} in general that for SSB drives there is no analogue of the `pumped charge' appearing at the edges, which gave rise to protected edge modes in the unbroken SPT ordered drives examined in I.

\subsubsection{Paramagnetic Phases}\label{ss:paramagnetic}
As an aside, we very briefly comment on the paramagnetic phases. The physics of these regions is covered in I, where we argued that the Floquet paramagnetically ordered drives with unitary finite symmetry group $G$ are classified by $\text{Cl}_G\times \mathcal{A}_G$ where $\text{Cl}_G$ is the SPT classification for $G$ and $\mathcal{A}_G$ are the 1D reps of $G$. In the present case $G=\mathbb{Z}_2$, for which $\text{Cl}=\{0\}$ and $\mathcal{A}=\mathbb{Z}_2$ so that there are two qualitatively different Floquet drives with PM order. In contrast to the FM case it turns out that in line with I, both of these drives have the same bulk spectral properties, and the difference between them is only visible on a system with a boundary. We present examples of drives in these two classes in \appref{app:Ising}. 

Two additional observations may interest the reader. First, as noted in Ref.~\onlinecite{Khemani15}, the PM phases are related to the FM phases by duality and hence can be diagnosed by bulk dual order parameter correlations which are non-local in the spin variables. Second, the two PM and two FM phases are related by Jordan-Wigner transformation to the class D drives examined in Ref.~\onlinecite{vonKeyserlingkSondhi16a} -- the FM regions correspond to those drives with bulk SPT (Kitaev wire-like order), while the PM regions correspond to the drives with trivial (non-SPT) bulk order. 

\subsection{Generalizing to the MBL regime} \label{s:towards}
%\section{Towards a more general picture}
We have examined two idealized Floquet drives (FM and $\pi$ FM) with Ising ferromagnetic order, and shown that they have markedly different spectral structure. The goal of this section is to argue, based on assumptions to be stated, that the differences between these Floquet drives is robust to adding interactions and strong disorder -- in other words to show that the FM and $\pi$ FM drives are representatives of two sharply defined phases. We then attempt to distill the previous section's observations into a more readily generalizable framework, which we apply to drives with spontaneously broken $\mathbb{Z}_n$ eigenstate order in \secref{ss:Zn}. Building on this in \secref{s:nonab} we  present the general case of Floquet drives with spontaneously broken non-abelian symmetry. 

Recall that the idealized drives in \secref{s:IsingPhaseDiagram} have Floquet unitaries with a set of exactly local bulk l-bits of form $Z_s Z_{s+1}$. How might this picture change away from the ideal point, in the presence of strong disorder and interactions? Returning to the discussion of the Ising ordered drives in \secref{ss:FMregions}, recall that the Ising order parameter $Z_r$ operators can be used to toggle between the degenerate (or $\pi$ quasi-energy split) Ising even/odd eigenstates. Using the specialized drives \eqnref{eq:classDUf} these degeneracies (or $\pi$ quasi-energy pairings) are exact. Perturbing symmetrically away from the fixed point, and in the presence of sufficiently strong disorder, we expect the degeneracies (or $\pi$ quasi-energy pairings)  to be exact up to exponentially small corrections in system size\cite{Huse13}. In line with the expected behavior of MBL phases in the undriven setting\cite{Huse13,Huse14}, and the observed behavior of the Floquet spectra in numerics\cite{Khemani15}, we will henceforth assume that upon perturbing symmetrically away from the fixed point the SSB ordered Floquet drives obey the following conditions: There exist smeared out (but local) analogues of $Z_s$ (which we continue to denote $Z_s$) which are Ising odd, obey $Z^2_s=1$, and which commute amongst themselves all up to exponentially small corrections in the system size. In particular, the operators $Z_s Z_{s+1}$ continue to be Ising even, and are the l-bits of the new system, also up to exponentially small corrections in the system size. The upshot is that we are working now with l-bits and order parameters very like those in the idealized models, but many of the previous relations between these operators hold only up to exponentially small corrections in system size, which we henceforth ignore.

With these assumptions in place, we can constrain the general form of an Ising symmetric Floquet unitary $U_f$ with full SSB eigenstate order. Such
a Floquet unitary has a set of l-bits $B_{s} = Z_s Z_{s+1}$
only $N-1$ of which are independent. Note however that
$ P, B_{s} $ gives a set of $N$ independent integrals of the motion. By this we mean there is a complete eigenbasis labelled by $P=p,B_s =b_s$
\be\label{eq:Ufonlbitbasis}
U_{f}\mid p,\{b_{s}\}\rangle=u_{f}(p,\left\{ b_{s}\right\} )\mid p,\{b_{s}\}\rangle
\ee
where the eigenvalues $u_f$ depend on the $p,b_s$ eigenvalues. It follows straightforwardly that we can write $U_{f}$ entirely in terms of operators
$B_{s},P$
\be\label{eq:PBs}
U_{f}=U_{f}(P,\{B_{s}\})\punc{.}
\ee
We now use the locality of the instantaneous drive Hamiltonians $H(t')$ to constrain the functional dependence on $P$. To begin, note that $U_f$ commutes with $Z_l Z_{r}$ for any $l,r$. In other words, using notation $[A:B]=A B A^{-1} B^{-1}$ for unitaries $A,B$ we have
\be\label{eq:commutatorIsing}
[U_f:Z_l Z_r]=1
\ee
As $U_f$ is a local unitary, $U_f Z_l U^{-1}_f =Z_l \theta_l$ and $U_f Z_l U^{-1}_f = \theta_r Z_r $, where $\theta_{l,r}$ are unitaries with exponentially localized support near $l,r$ respectively using Lieb-Robinson bounds\cite{Lieb72}. However, \eqnref{eq:commutatorIsing} implies that $\theta_l = \theta^{-1}_r$, for $l,r$ arbitrarily distant from one another. The only possible resolution is that $\theta_l= \theta^{-1}_r = e^{i\varphi}1$ for some phase $e^{i\varphi}$. This phase is moreover constrained to be $e^{i\varphi}=\pm1$, using $Z^2_l=1$ and the resulting equality $1= U_f Z^2_l U^{-1}_f = \theta^2_l 1$. It follows that
\begin{align}
Z_l U_f (P,\{ B_s\} ) Z^{-1}_l &= U_f (-P,\{ B_s\} ) \nonumber\\
& =\pm U_f (P,\{ B_s\} )\punc{,}
\end{align}
where the second equality reflects the conclusion $e^{i\varphi}=\pm1$ in the paragraph above, while the first equality follows from the fact that $Z_l$ anti-commutes with global Ising generator $P$. In the $+1$ case $U_f$ is independent of $P$, while in the $-1$ case it is odd in $P$, i.e.,
$$
U_f= U'_f (\{B_s\})  \text{ or } P U'_f (\{B_s\}) 
$$
for $e^{i\theta}=\pm1$ respectively, where $U'_f (\{B_s\}) $ is some unitary depending only on the bond operators $B_s$. These two distinct types of Floquet unitaries are consistent with the different structures of the FM and $0\pi$ FM examined in the examples in \secref{ss:FMregions}.  Indeed, by the same reasoning as in \secref{ss:FMregions}, they have the same spectral properties. Namely, labelling the eigenstates of $U_f,P$ by pairs $(u,p)$: In the $e^{i\varphi}=1$ case there are protected doublets of eigenstates with $(u,+1),(u,-1)$, while in the $e^{i\varphi}=-1$ example there is a $\pi$ quasi-energy spectral pairing ,i.e., doublets of states with $( u,1),(-u,-1)$.

In summary we have argued that there are two fundamentally different kinds of Ising symmetric Floquet unitary with FM Ising order, distinguished by their commutation with the Ising order parameter ${[U_f  :Z_l]=\pm 1}$. The FM and $0\pi$ FM Floquet drives examined in \secref{s:IsingPhaseDiagram} are idealized examples of these two kinds of unitary.  We have not yet argued, however, that these distinct kinds of Floquet unitary define genuinely distinct Floquet phases stable to sufficiently small perturbations to the unitary. We will return to this issue when we treat the general case, but we give a summary of the argument here. 

We now argue that the eigenstate properties of the Floquet drives constructed above are stable to sufficiently small Ising symmetric changes to the Floquet drive. Assuming a small change in $U_f$ leads to a small change in the $Z_l$ operators, then $[U_f:Z_l]$ must change by a small amount as well. However, we have argued that this quantity is independent of $l$  and discrete --  above we argue it is equal to $\pm 1$. Hence it cannot change continuously, so it does not change at all. In this way, making the stated assumptions about the forms of the l-bits and their dependence on $U_f$, we have argued that our diagnostic $[U_f:Z_l]=\pm 1$ distinguishing different Floquet drives is robust. Hence we expect the two distinct drives constructed in \secref{s:IsingPhaseDiagram} correspond to genuinely distinct Floquet phases. Having discussed the Ising case in general we now examine more briefly how these results generalize to Floquet drives with $\mathbb{Z}_n$ completely SSB order.

\section{$\mathbb{Z}_n$ SSB drives}\label{ss:Zn}
Here we extend the results in the previous section to theories with completely broken $\mathbb{Z}_n$ symmetry, and introduce a notation which more readily generalizes to the non-abelian cases studied in \secref{s:nonab}. In the Ising case our on-site Hilbert space consists of $ Z = \pm 1$ on-site Ising degrees of freedom. In the rest of the paper, we consider the more general Hilbert space $\mathcal{H}$, where the on-site Hilbert space consists of $ g_r \in G$ degrees of freedom, where $G$ is an on-site unitary global symmetry group.

It is convenient to view $G=\mathbb{Z}_n$ as a subset of U$(1)$ generated by $\omega = e^{2\pi i/n}$. Let us now define some useful linear operators (which live in $\mathcal{L}(\mathcal{H})$). First, there are the global $\mathbb{Z}_n$ symmetry generators $V:G\rightarrow \mathcal{L}(\mathcal{H})$ of form $V(g)=\prod_r V_r(g)$ where
\be\label{eq:onsite}
V_r(x) \mid  \{g_s\} \rangle = \mid   \{x^{\delta_{r s}} g_s\} \rangle
\ee
for any $x,g_r\in G$. In the Ising problem $V_r(\pm 1)=1,X_r$ respectively, and $V(\pm 1)=1,P$ respectively. Additionally, define a unitary operator
$$
\mfg_r \mid \{g_s\} \rangle = g_r \mid  \{g_s\} \rangle\punc{,}
$$
where $g_r \in G$ is taken to be an $n^{\text{th}}$ root of unity. In the $G=\mathbb{Z}_2$ case, $\mfg_r=Z_r$. The commutator
\be
[V_r(x):\mfg_r]=x^{-1} \in \text{U}(1) \punc{,}
\ee
follows from these definitions, as does
\be
[V(x):\mfg_r]=x^{-1} \in \text{U}(1)\punc{.}\label{eq:commZn}
\ee
At this point, let us describe more precisely what we mean by SSB eigenstate order. It is useful to give an example of a fixed point Hamiltonian with SSB $\mathbb{Z}_n$ order 
\be\label{eq:ZnSSB}
H_1 = \sum_r  J_r(\mfg^{\dagger}_r \mfg_{r+1})\punc{,}
\ee
where $J_r$ is some disordered Hermitian function of the $\mathbb{Z}_n$ variables. Note that $H_1$ commutes with the global symmetry generators $V(x)$ using \eqnref{eq:commZn}.  We interpret $\mfg_{r}$ as our new $\mathbb{Z}_n$ valued order parameter, and as in the Ising case we will assume that away from the fixed point model Hamiltonian \eqnref{eq:ZnSSB}, there are smeared out analogues of $\mfg_{r}$ which commute amongst themselves, obey \eqnref{eq:commZn} with the global symmetry generators $V(x)$, as well as $\mfg^n_{r}=1$, all up to exponentially small corrections in system size which we ignore.   As a corollary, $\mfB_r\equiv\mfg^{\dagger}_r \mfg_{r+1}$ are a set of l-bits. These we take to be the defining features of $\mathbb{Z}_n$ eigenstate order.  

We now investigate unitaries $U_f$ with the aforementioned eigenstate order -- i.e., with a set of $\mfg_r$ operators, and a set of local integrals of motion $\mfB_r=\mfg^{-1}_r \mfg_{r+1}$. The variables $\mfB_r , V(\omega) $ constitute a complete set of conserved quantities which all commute with the global symmetry generators -- there are $N-1$ independent l-bits, taking $n$ possible values, and the global symmetry generator $V(\omega)$ taking $n$ possible values. The set is complete because the total degrees of freedom $N^n$ coincides with the total Hilbert space size. Note that we  call the $\mfB_r= \mfg^{\dagger}_r \mfg_{r+1}$ variables l-\textit{bits}, even though they take values in the n$^{\text{th}}$ roots of unity. Just as was argued in the Ising case (near \eqnref{eq:Ufonlbitbasis}) the unitary must have functional dependence
$$
U_f =U_f (\{\mfg^{\dagger}_s \mfg_{s+1}\}, V(\omega)) \punc{.}
$$
Using a straightforward extension of the argument below \eqnref{eq:PBs} where we showed $[U_f:Z_l]=\pm 1$, it follows that $[U_f:\mfg_l]= \omega^{-k}\in \mathbb{Z}_n$ for some $k$. This in turn ensures that the Floquet unitary takes the form
\be\label{eq:Zncanonicalform}
U_f =U_f (\{\mfg^{\dagger}_s \mfg_{s+1}\})  V(\omega^{k}) \punc{,}
\ee
for some $k=0,1,\ldots,n-1$. In fact, we can readily construct any such unitary taking local Hamiltonian $H_0 = \sum_s \log(V_s(\omega^{k}))$ and combining it with $H_1$ defined in \eqnref{eq:ZnSSB} according to the prescription \eqnref{eq:classDUf}. We are thus led to the conclusion that there are $n$ distinct $\mathbb{Z}_n$ completely symmetry broken drives.

Let us briefly mention the kinds of spectral pairing in this model. Fixing $k$ in \eqnref{eq:Zncanonicalform}, we find the Floquet spectrum has degenerate multiplet of $n$ states at each quasi-energy. This follows from choosing a simultaneous eigenbasis of $U_f,V(\omega)$ labelled by $(u,v)$. Applying the operators $1,\mfg_l,\ldots,\mfg^{n-1}_l$ to this state generates a multiplet of states $(u,v ),(u\omega^{-k},v\omega^{-1}),\ldots,(u\omega^{-k(n-1)},v\omega^{-(n-1)})$. So, for $k=0$ the eigenstate order is simply that of a SSB undriven $\mathbb{Z}_n$ state, exhibiting the characteristic $n$-fold degeneracy. For $k\neq 0 \mod n$ there is no undriven analogue: we find a multiplet of $n$ states with protected quasi-energy gaps. We can also argue that no protected edge modes arise when we restrict such a unitary to a system with edges -- see \secref{ss:nothingnewattheedge}.

Last we argue that the eigenstate properties of the newly predicted Floquet drives constructed above are stable to small symmetric perturbations. If we assume that the operators $\mfg_r$ change continuously as we perturb $U_f$ symmetrically, then $[U_f:\mfg_r]$ changes continuously too --  the analogue of this quantity in the Ising case was $[U_f:Z_r]$. However, we have argued that this quantity is a pure phase and discrete -- it is an $n^{\text{th}}$ root of unity. Hence, it cannot change continuously. So it does not change at all. In this way, with some assumptions about how the l-bits vary with small changes in $U_f$, we have argued that our diagnostic distinguishing different Floquet drives is robust. Hence we expect that the $n$ types of Floquet unitaries -- listed in, and constructed explicitly below \eqnref{eq:Zncanonicalform} -- correspond to distinct and stable Floquet phases. We have extended many of the arguments of \secref{s:towards} to drives with $\mathbb{Z}_n$ SSB eigenstate order, paving the way for the case of general finite $G$.

\section{General finite $G$ SSB order}\label{s:nonab}
Here we classify Floquet unitaries with SSB eigenstate order, for general potentially non-abelian $G$. This section is structured as follows. After setting up some notation, we state more comprehensively what we mean by SSB eigenstate order for general $G$. Using this definition, and certain more technical locality arguments in \secref{app:notlowdepth}, we constrain the form of a Floquet unitary to \eqnref{eq:Ufcanonical}, showing that the different classes of fully SSB ordered Floquet unitaries $U_f$ are labelled by the elements of $Z(G)$, the center of the group $G$. Thus, we  predict a $Z(G)$ classification for completely spontaneously broken Floquet drives. Last in \secref{ss:nothingnewattheedge} we argue that there are no protected edge modes for the predicted fully SSB Floquet phases. 

Consider a Hilbert space  with on-site $g_r \in G$ degrees of freedom, with $G$ potentially non-abelian.  It is useful to identify each such $g_r$ with its matrix $\{g_{r,ij}\}$ taken in some complex faithful representation of $G$ (e.g., the regular representation always works) so that in particular
\be\label{eq:groupcomp}
\sum^{d}_{k=1} g_{ik} h_{kj} = (gh)_{i j}\punc{,}
\ee
where $d$ is the dimension of the representation. As in \eqnref{eq:onsite} we define on-site symmetry generator $V_r(x) {\mid \{g_s\} \rangle = \mid \{x^{\delta_{rs}} g_s\} \rangle}$, and right multiplication $V^{\text{op}}_r(x) {\mid \{g_s\} \rangle = \mid \{g_s x^{\delta_{rs}} \} \rangle}$. As for the $\mathbb{Z}_n$ case, we define `order parameter' operators $\mfg_r$, except these operators are matrix valued with
\be\label{eq:defmfg}
\mfg_{r,ij}  \mid \{g_s\} \rangle = g_{r,ij}\mid  \{g_s\} \rangle 
\ee
where $i,j =1,\ldots,d$. It follows from \eqnref{eq:groupcomp} that $\mfg_r$ is a unitary matrix in the sense that
$$
\sum^d_{k=1} \mfg^{\dagger}_{r,k i } \mfg_{r,kj}  = \delta_{i j} 1\punc{.}
$$
Similarly, we obtain commutation relations
\begin{align}\label{eq:nonabcomm}
V_r(x) \mfg_{r,ij} V^{-1}_r(x) &= \sum_k (x_r)^{-1}_{ik} \mfg_{r,kj} \text{, and}\\
V(x) \mfg_{r,ij} V^{-1}(x) &= \sum_k (x_r)^{-1}_{ik} \mfg_{r,kj}\label{eq:nonabcommglobal}
\punc{.}
\end{align}
To describe more precisely what we mean by SSB eigenstate order for non-abelian $G$, it is useful examine the fixed point Hamiltonian
\be\label{eq:nonabferro}
H_1 = \sum_r J_r ( \mfg^{\dagger}_r \mfg_{r+1}  )\punc{,}
\ee
where the matrix indices on $\mfg_r$ are left implicit.  Here $J_r:G\rightarrow \mathbb{R}$ is some random set of functions associated with the $r,r+1$ bond. \eqnref{eq:nonabferro} is our prototypical example of SSB for general finite group $G$.  Using \eqnref{eq:nonabcomm}, $H_1$ is $G$ symmetric.  Note too that 
\be\label{eq:mfB}
\mfB_r \equiv \mfg^{\dagger}_{r} \mfg_{r+1}
\ee
is a near complete set of local conserved quantities, which entirely determine the positions of domain walls in a `spin-glass' configuration of a group valued order parameter $\mfg_r$ -- specifically, these operators tell us how the order parameter changes as we hop across the bond. Now pick a particular site $r=1$. The operators $\mfB_r , \mfg_1 $ give a complete set of labels on the whole Hilbert space. That is, a spin glass configuration is completely specified by the order parameter at a particular site $\mfg_1=g_1$ and the manner in which the order parameter changes site to site $\mfB_r=B_r$.  In analogy with the results in the previous section, perturbing symmetrically away from the fixed point model, we expect a smeared out analogue of the $\mfg_r$ operators obeying commutation relation \eqnref{eq:nonabcommglobal} with the global symmetry generator $V(x)$. Additionally, we expect modified l-bits of form $\mfB_r = \mfg^{\dagger}_{r} \mfg_{r+1}$, with $\mfB_r=B_r,\mfg_1=g_1$ giving a complete set of conserved quantities (all commuting up to exponentially small corrections in system size).

Having defined SSB order, we move to the Floquet problem. Consider a local symmetric unitary  $U_f$ which has SSB ordered eigenstates as per the above specification. The unitary $U_f$ must commute with all of the l-bits $\mfB_r$, implying
$$
U_f = \sum_{g,g'} \mfu'_{\{B\}}(g',g) \mid g', \{B\} \rangle \langle g, \{B\}  \mid
$$
where $g,g'$ is the value of the group element at site $1$  and we have chosen to label our basis states by ${\mfg_1=g_1,\mfB_r=B_r}$. We are interested only in some such operators which commute with the global symmetry generators $V(x)$. This imposes condition $\mfu'_{\{B\}}(g',g)=\mfu'_{\{B\}}(x g',x g)$ for any $x$, which is equivalent to the statement that
\begin{align}
U_f&= \sum_{g,g'} \mfu_{\{B\}}(g^{-1}  g' ) \mid g', \{B\} \rangle \langle g, \{B\}  \mid\nonumber \\
&= \sum_{g,g'} \sum_x \delta(x = g^{-1}  g') \mfu_{\{B\}}( x ) \mid g x , \{B\} \rangle \langle g, \{B\}  \mid\nonumber\\
&= \sum_{x} \mfu_{\{B\}}( x ) \underbrace{\sum_g \mid g x , \{B\} \rangle \langle g, \{B\}  \mid}_{Q(x)}\punc{.} \label{eq:Q}
\end{align}
where $\mfu'_{\{B\}}(g',g) = \mfu_{\{B\}}(g^{-1}g')$ defines $\mfu$. In the original convention for labelling basis vectors with their $\mfg_r$ eigenvalues, this newly defined operator $Q$ acts like
\be\label{eq:Qredefined}
Q(x)\mid\{g_r\}\rangle =\mid\{g_1 x g^{-1}_1 g_r \}\rangle\punc{.}
\ee
For $x\in Z(G)$ the centre of $G$, $Q(x)$ just acts like the global left symmetry action $V(x)$, and in particular $Q(x)$ is a local circuit. For non-central $x$ this operator is not low depth (see \appref{app:notlowdepth}). Moreover we show in \appref{app:notlowdepth} that $U_f$ is local only if $\mfu_{\{B\}}( x )$ vanishes except for $x=z$, where $z$ is a particular element in the center. In other words
\be
U_f =   \mfu_{\{\mfB_r\}} V(z)  \label{eq:Ufcanonical}\punc{,}
\ee
where $\mfu_{\{\mfB_r\}}$ is some unitary function of the l-bits and $z\in Z(G)$ is a particular element of the center. We thus arrive at the conclusion that Floquet unitaries with full SSB order are characterized by some $z\in Z(G)$. Using the methods of I, $\mfu_{\{\mfB_r\}}$ can be argued to be a local functional of domain wall configurations. 

Last we might ask if there is an operator diagnostic allowing us to discern the value of $z$ appearing in \eqnref{eq:Ufcanonical}. Indeed there is.  Let $g^{\chi}_{ij}$ be a matrix presentation of $g\in G$ within irreducible representation $\chi$ of $G$, with $i,j\in 1,\ldots, \text{dim}_\chi$.  Define, in analogy with \eqnref{eq:defmfg},
\be\label{eq:mfgchi}
\mfg^{\chi}_{r,ij}  \mid g_r \rangle = g^{\chi}_{r,ij}\mid g_r \rangle \punc{.}
\ee
Using \eqnref{eq:nonabcomm}, and the fact that $z\in Z(G)$ acts like a phase in any irreducible representation, it follows that
\be\label{eq:mfgchicomm}
V(z)  \mfg^{\chi}_{r,ij}V^{-1}(z) = \frac{\chi(z)}{\chi(1)}\mfg^{\chi}_{r,ij}\punc{,}
\ee
where $\chi$ is the irreducible character, and $\chi(z)/\chi(1)$ is a pure phase again because $z\in Z(G)$. Multiplying \eqnref{eq:mfgchicomm} by $\mfg^{\dagger}_{r,j 1}$, setting $i=1$, and summing over $j$ gives
$$
\sum_j V(z)  \mfg^{\chi}_{r,1j}V^{-1}(z) \mfg^{\dagger,\chi}_{r,j 1}= \frac{\chi(z)}{\chi(1)}\punc{.}
$$
Now using the orthogonality of the character table 
\be\label{eq:orthog}
\sum_\chi \chi^{*}(C) \chi(C') = \delta_{C,C'}\frac{|G|}{|C|}
\ee
for conjugacy classes $C,C'$ we find that 
for any $z'\in Z(G)$
$$
 \frac{1}{|G|} \sum_{j,\chi} V(z)  \mfg^{\chi}_{r,1j}V^{-1}(z) \mfg^{\dagger,\chi}_{r,j 1}\chi^{*}(z')= \delta_{z,z'}\punc{.}
$$
Using this identity, and the fact that $\mfg^\chi$ commute with the l-bits, we extract $z$ from $U_f$ in \eqnref{eq:Ufcanonical} using operation
\be \label{eq:determinez}
 \frac{1}{|G|} \sum_{j,\chi} U_f  \mfg^{\chi}_{r,1j}U^{-1}_f \mfg^{\dagger,\chi}_{r,j 1}\chi^{*}(z')= \delta_{z,z'}\punc{.}
\ee
Having argued that $U_f$ takes canonical form \eqnref{eq:Ufcanonical} for some $z\in Z(G)$, we now argue that the  Floquet drives corresponding to different $z$ correspond to distinct stable Floquet phases. We argue for the stability of these phases much as we did in \secref{ss:Zn}. In the present case the quantity \eqnref{eq:determinez} for any $z'\in Z(G)$ entirely determines the $z\in Z(G)$ characterizing the Floquet unitary \eqnref{eq:Ufcanonical}. It is discrete (either $0,1$), and appears to depend continuously on $U_f$, so by the argument in \secref{ss:Zn} it is expected to be stable to sufficiently small symmetric perturbations. 

\subsection{The absence protected edge modes in SSB ordered drives}\label{ss:nothingnewattheedge}

Having classified completely SSB Floquet phases according to their bulk spectra, we now argue that they have no protected edge modes -- this is in contrast to the unbroken examples in I, where the non-triviality of the Floquet drives manifested itself through the presence of additional (or modified) edge modes. Consider an SSB drive on a system with boundary. Using arguments like those in I  (App.'s A and B), together with  \eqnref{eq:Ufcanonical}, we can argue that the Floquet unitary on a system with large number $N$ sites takes the form
\be\label{eq:decompold}
U_f = v_L v_R e^{-i f}V(z)\punc{,}
\ee
where $z\in Z(G)$, $f$ is a local function only of bulk l-bits, and $v_L,v_R$ are unitaries local to the $L,R$ edge of the system which commute with all the bulk l-bits . As in I, we can show that $v_L,v_R$ commute with global symmetry $V(g)$ up to some phase characterized by a pumped charge $[V(g):v_L]=\kappa(g)$ for some 1D representation $\kappa$ of $G$. However it turns out that due to the bulk SSB order, the pumped charge is just an artifact of the particular way we have decomposed the unitary in \eqnref{eq:decompold}, rather than a robust feature of the unitary. 

To see why, form operators $\mfg^{\kappa}_r$ corresponding to the 1D representation $\kappa$. As $\kappa$ is a 1D representation, $\mfg^{\kappa}_r$ is a scalar unitary operator with eigenvalues which are roots of unity (as opposed to a matrix valued operator like $\mfg_{r,ij}$).  Now redefine  $v_{L,R}$ by multiplying them with $\mfg^{\kappa}$ operators based at the left/right of the system respectively: $v'_{L,R}=v_L \mfg^{\dagger,\kappa}_L, v_R \mfg^{\kappa}_R$.  Simultaneously, redefine $f$  by adding a local Hermitian functional of the l-bits $f'=f+i\sum^{R-1}_{s=L} \log (\mfg^{\kappa}_s \mfg^{\dagger,\kappa}_{s+1})$. Using the identity
\be
e^{\sum^{R-1}_{s=L} \log (\mfg^{\kappa}_s \mfg^{\dagger,\kappa}_{s+1}) } = \prod^{R-1}_{s=L}  \mfg^{\kappa}_s \mfg^{\dagger,\kappa}_{s+1}å\propto \mfg^{\kappa}_L \mfg^{\dagger,\kappa}_{R}\punc{,}
\ee
together these modifications leave $U_f$ unchanged, but $v'_{L,R}$ now commute with the global symmetries.  Therefore we may as well assume $v_L,v_R$ are symmetric. This means we can exchange them for any local symmetric edge unitaries while preserving the symmetry of $U_f$. In particular, there are no protected edge states. The physical intuition behind this calculation is as follows. In the dual language, we can view SSB order as a condensate of particles carrying representations of $G$ -- e.g., in dual variables, the Ising ferromagnet is a condensate of Ising odd particles. In such a situation, there is no solid notion of pumped charge --  any charge pumped into the edge is immediately screened by the delocalized soup of charges in the bulk.

\section{Constructing drives}\label{s:nonabexamples}
The explicit Ising symmetric drives we examined at the start of this work inspired a more general classification of SSB Floquet phases. We saw that Floquet MBL phases with a fully spontaneously broken finite symmetry group $G$ are classified by $Z(G)$. We now construct explicit (fixed point) drives for each of the new predicted phases. Pick a $z\in Z(G)$ labelling the desired phase and let
\be
K_z =- \sum_r i \log(V_r(z))\punc{.}
\ee
As $V_r(z)$ is a unitary operator, a logarithm $\log V_r(z)$ exists -- for concreteness, we define this explicitly in \eqnref{eq:log}. Now take the random SSB spin-glass Hamiltonian $H_1$ from \eqnref{eq:nonabferro}. The unitary circuit
\be
U(t)=\begin{cases}
e^{-iK_z t} & 0 \leq t<1  \\
e^{-iH_1 (t-1)}e^{-iK_z} &1\leq t<1+ t_{1} 
\end{cases}
\ee
has Floquet unitary
$$
U_f = e^{- i H_1 t_1} V(z) \punc{,}
$$
where $V(z)$ is a global symmetry generator, and $H_1$ has a fixed point spin-glass order and is a functional only of the $\mfB_r$ operators from \eqnref{eq:mfB}. This unitary explicitly of the form \eqnref{eq:Ufcanonical},  for the phase corresponding to $z\in Z(G)$.

\section{Structure of the Floquet spectrum and time crystals}\label{s:algebraic}
The Floquet phases predicted above are characterized by a central element.  The spectral properties of these drives are obtained by considering a subspace corresponding to some fixed configuration of domain walls $\{B\}$. We wish to consider the possible values of $U_f$ on this subspace. We can certainly decompose this subspace into irreducible representations of $G$. Starting in a state which is a singlet under global symmetry, we  can toggle  to a state living in irreducible representation $\chi$ using operator $\mfg^\chi_{r,i j}$. However,  $U_f \mfg^{\chi}_{i j} U^{-1}_f= \frac{\chi(z)}{\chi(1)} \mfg^{\chi}_{i j}$ where $\chi(z)$ is the character evaluated at $z\in Z(G)$ corresponding to irreducible representation $\chi$. In other words, the Floquet evolution flips our non-abelian generalization of a spin glass order parameter $\mfg^{\chi}_{i j}$. We see that states living in irreducible representation $\chi$ have their spectra shifted by $\chi(z)/\chi(1)$ relative to the original state -- this is a pure phase because $\chi$ is an irreducible representation and $z\in Z(G)$\footnote{$z$ gives rise to a unitary automorphism of irreducible representation space corresponding to $\chi$, so by Schur's lemma acts like a phase $e^{i \theta} 1$.}. 

With the basic structure of the spectra in place, we note a connection between our work and time crystals\cite{Wilczek12}. The Hamiltonians $H(t)$  for the Floquet phases considered above not only have an on-site symmetry group $G$, but also have a symmetry under time translation $H(t+T)=H(t)$. In this sense, the total symmetry group is $G\times \mathbb{Z}$, where $\mathbb{Z}$ represents time translation.  As stated above the additional information characterizing the drives is an element of the center $z$, or equivalently a homomorphism from the abelian group of time translations to the global symmetry group $\varphi:\mathbb{Z} \rightarrow G$.

The drives above spontaneously and completely break the symmetry $G$, but there is also a sense in which they spontaneously break the Floquet time translation symmetry $t\rightarrow t+T$, in a manner characterized by the central element $z$ alluded to above. For $z\neq 1$ the order parameter  oscillates non-trivially
$$
\mfg^{\chi}_{i j}(nT)=\left[\frac{\chi(z)}{\chi(1)}\right]^n \mfg^{\chi}_{i j}(0)\punc{,}
$$ 
with period larger than $T$, even though  the Hamiltonian has period T. In other words, the order parameter time dependence does not enjoy $t\rightarrow t+T$ translation symmetry. In the $\pi$FM Ising case of \secref{s:IsingPhaseDiagram}, this is the statement that the order parameter oscillates $Z(nT) = (-1)^n Z(0)$ with period $2T$ oven though the underlying Hamiltonian has period $T$. The connection between this statement, and time crystals was pointed out to us by E. Altman\footnote{Private communication.}. This oscillation is detectable if one prepares a spin-glass configuration state and measures the SG order parameter stroboscopically, although the full time dependence of spins in such systems is much more complicated as we will discuss elsewhere\footnote{C.W. von Keyserlingk, V. Khemani and S.L. Sondhi, in preparation.}. This notion of time crystal is close in spirit to the attempted definition in Ref.~\onlinecite{Oshikawa15}, where a no-go theorem was proved concerning spontaneous breaking of continuous time translation symmetry. This no-go theorem is inapplicable to the present systems for a number of reasons, in particular our Hamiltonians explicitly break continuous time evolution symmetry.

\section{Generalizations and conclusion}\label{s:conclusion}
We have put forward a classification scheme for 1d many-body localized Floquet SPT states with completely spontaneously broken on-site symmetry $G$, and with on-site group valued degrees of freedom. We conjecture that there are $|Z(G)|$ different possible Floquet drives, each of which can be brought into a canonical form \eqnref{eq:Ufcanonical}. We have argued that these putative Floquet phases are stable to sufficiently small modifications to the unitary $U_f$ in the bulk, although our arguments are only heuristic and make certain assumptions about the behavior of l-bits away from our exactly solvable fixed points.  

The current work can be extended in several directions. Although we have focussed on 1d, none of the arguments seem specific to 1d, so we tentatively conjecture a $Z(G)$ classification for higher dimensional completely symmetry broken phases too. However, with the nature and stability of MBL order in higher dimensions currently in question, we make this proposal very tentatively. As in our previous work I, there remains the challenge of understanding the dynamical stability of these new phases for realistic drives, and the need for proposals for realizing and detecting them in experiments.

\acknowledgements
We thank V. Khemani, R. Moessner and A. Lazarides for many discussions and for collaboration (with SLS) on prior work. We are grateful to E. Altman for suggesting a connection between our work and time crystals. CVK is supported by the Princeton Center for Theoretical Science. SLS would like to acknowledge support from the NSF-DMR via Grant No.~1311781 and the Alexander von Humboldt Foundation for support during a stay at MPI-PKS where this work was begun.
\begin{appendix}

\section{Ising paramagnetic regions}\label{app:Ising}
In this section, we investigate binary drives of form \eqnref{eq:classDUf} corresponding to the PM regions in \figref{classDPhasediagram}. To distinguish the two possible Floquet unitaries, we will need to consider the drives on a system with boundary. We will here demonstrate the existence of two distinct such paramagnetic Floquet drives by looking at specific points on the \figref{classDPhasediagram} phase diagram. The two distinct drives correspond to two possible phases of the $\text{Cl}_G\times\mathcal{A}_G=\mathbb{Z}_2$ classification for $G=\mathbb{Z}_2$ in I.

In the region labelled PM in \figref{classDPhasediagram}, all of the eigenstates have paramagnetic order. A representative unitary is obtained by setting $t_1=0$ i.e., $U_f=e^{-i H_0 t_0}$, in which case that the eigenstate properties of this unitary are simply those of the topological hamiltonian $H_0$ with l-bits of form $X_s$. Note that a such a PM hamiltonian (in the disordered setting) does not have the spectral pairing present in the FM problem. 

Finally consider the $0\pi \text{ (PM)}$ on an open system. For ease of explanation set $J_s=1$ and $h_s$ disordered. As an example, set $t_0  < \frac{\pi}{2}$ and $t_1=\pi/2$.
\begin{align}
U_f &=\prod^{N-1}_{s=1}Z_s Z_{s+1} e^{-  i t_0 H_0}  \rightarrow  Z_1 Z_N e^{-  i t_0 \sum^{N-1}_{s=2} h_s X_s} 
\end{align}
where  we performed a local symmetric unitary change of basis to simplify the unitary near the edges. Now in the bulk ($s=2,\ldots,N-1$), the $U_f$ eigenstates are  eigenstates of  the local bulk integrals of motion $X_s$. In total  $U_f$ looks like a bulk PM drive multiplied by an Ising tunneling operator $Z_1 Z_N$. Note that the edge degrees of freedom are completely decoupled from the bulk so we can separately diagonalize the bulk hamiltonian  $e^{-  i t_0 \sum^{N-1}_{s=2} X_s} $ and the two site unitary 
 $$
 U_{f,\text{edge}} = Z_1 Z_N
 $$
 This two-site Hamiltonian has two useful independent integrals of motion $ U_{f,\text{edge}} =Z_1 Z_N$  and $ P_{\text{edge}} =X_1 X_N$ -- note these are also integrals of motion of the original unitary $U_f$. Picking a reference eigenstate $\mid 1,1\rangle$, we can toggle between the four eigenstates of $U_{f,\text{edge}}$.

\begin{table}[h]
\begin{tabular}{|c|c|c|}
\hline 
 & $U_{f,\text{edge}}$ & $P_{\text{edge}}$\tabularnewline
\hline 
$\left|1,1\right\rangle $ & $1$ & $1$\tabularnewline
\hline 
$X_1 \left|1,1\right\rangle $ & $-1$ & $1$\tabularnewline
\hline 
$ Z_1 \left|1,1\right\rangle  $ & $1$ & $-1$\tabularnewline
\hline 
$X_1 Z_1 \left|1,1\right\rangle  $ & $-1$ & $-1$\tabularnewline
\hline 
\end{tabular}
\caption{This table shows the structure of the spectrum of an Ising symmetric Floquet drive with paramagnetic order. }\label{tab:Z2representationtheory}
\end{table}%

Combining these edge results with the bulk unitary, we conclude that for the drive in question, eigenstates of the full unitary $\left|u,p\right\rangle$ of the Floquet drive come in quadruplets with $U_f,P$ eigenvalues $(u,p),(u,-p),(-u,p),(-u,-p)$.
\section{Locality arguments}\label{app:notlowdepth}
In this section we assume $U_f$ is local, and has a full set of l-bits of the form explained in \secref{s:nonab}. Using these assumptions we will show first that the operator $Q(x)$ defined in \eqnref{eq:Qredefined} is local iff $x\in Z(G)$. We then show that the full Floquet unitary takes the form $U_f = \mfu_{\{B\}}(z_0) V(z_0)$ where $z_0\in Z(G)$. The reader should beware that these two target equations, and many others in this section will hold only up to exponentially small corrections in system size.

\subsection{$Q(x)$ local iff $x\in Z(G)$}
We argue now that $Q(x)$ as defined above in \eqnref{eq:Q} is local iff $x\in Z(G)$. If $x$ is central, the conclusion follows readily from the fact that $Q(x)=V(x)=\prod_r V_r(x)$, which is manifestly local. If $x$ is not central, consider the operator $V_1(y)$ which has support on site $1$. Were $Q(x)$ local, a Lieb-Robinson bound would imply that $[Q(x):V_1(y)]$ commutes with operators based at sites $s$ very distant from $1$ (up to exponentially small corrections in $|s|$). Let $F$ be an operator $F\mid \{g_r\}\rangle\equiv \delta(g_s,1)\mid \{g_r\}\rangle$ -- clearly the operator has support only at site $s$. We will show that the commutator $[[Q(x):V_1(y)],F]$ does not decay with $s$. It suffices to show that some matrix elements of the commutator do not decay with $s$. Thus, we will have shown that $Q(x)$ is not a local unitary if $x\notin Z(G)$.

First, let us look at the matrix elements of ${[Q(x):V_1(y)]}$. For clarity we will calculate these step by step. First, recall

\begin{align}
Q(x) \mid \{ g_r \} \rangle &=  \mid  \{ g_1 x g^{-1}_1 g_r \} \rangle \nonumber \\
Q^{-1}(x) \mid \{ g_r \} \rangle &=  \mid  \{ g_1 x^{-1} g^{-1}_1 g_r \} \rangle \nonumber \\
\end{align}
so that $Q^{-1}(x)=Q(x^{-1})$. Next note that 
\begin{align}
&V^{-1}_1(y) \mid g_1, \{ g_r \}' \rangle \nonumber \\
&=  \mid y^{-1} g_1, \{ g_r \}' \rangle \nonumber \\
&Q^{-1}(x) V^{-1}_1(y) \mid g_1, \{ g_r \}' \rangle \nonumber \\
&=  \mid y^{-1} g_1 x^{-1}, \{  y^{-1} g_1 x^{-1}  g^{-1}_1 y  g_r \}' \rangle \nonumber \\
&V_1(y) Q^{-1}(x) V^{-1}_1(y) \mid g_1, \{ g_r \}'\rangle \nonumber \\
&=\mid  g_1 x^{-1}, \{  y^{-1} g_1 x^{-1}  g^{-1}_1 y  g_r \}' \rangle \nonumber \\
&Q(x)  V_1(y) Q^{-1}(x) V^{-1}_1(y) \mid g_1, \{ g_r \}' \rangle \nonumber \\
 &=\mid  g_1 , \{ g_1 x g^{-1}_1   y^{-1} g_1 x^{-1}  g^{-1}_1 y  g_r \}' \rangle\punc{.} \nonumber 
\end{align}
where the notation $ \mid g_1, \{ g_r \}' \rangle$ isolates the group element on site $1$ from all of the labels on other sites $\{g_r\}'=\{g_r:r\neq 1\}$. This allows us to calculate a group commutator between kets 
\begin{align}
&\langle \{g'_r \}\mid [[Q(x):V_1(y)],F]   \mid  \{g_r \} \rangle \nonumber \\
&= (F(g_s) - F(g_1 x g^{-1}_1   y^{-1} g_1 x^{-1}  g^{-1}_1 y  g_s ))\nonumber \\
&\times \delta_{g'_1,g_1} \delta_{g'_r, g_1 x g^{-1}_1   y^{-1} g_1 x^{-1}  g^{-1}_1 y  g_r } 
\end{align}
Using $F(g_s)=\delta_{g_s,1}$ and taking matrix elements betweens some state with $g'_1=g_1 = 1$, $g_s=1$ and $g'_r = g_1 x g^{-1}_1   y^{-1} g_1 x^{-1}  g^{-1}_1 y  g_r$ for $r>1$, we get 
\begin{align}
&\langle \{g'_r \}\mid [[Q(x):V_1(y)],F]   \mid  \{g_r \} \rangle \nonumber \\
&= (1 - \delta_{ x  y^{-1}  x^{-1}  y,1})\nonumber
\end{align}
This latter expression is equal to $1$ provided we can find a $y$ which fails to commute with $x$. This statement is true regardless of have large we choose $s$. Hence the operator norm of the commutator does not decrease exponentially with $s$.
\subsection{$U_f$ local only if $\mfu_{\{B\}}(x) \propto \mfu_{\{B\}}(z_0) \delta_{x,z_0}$ where $z_0\in Z(G)$ }
Given a \textit{local} unitary of form
\be
U_f= \sum_{x} \mfu_{\{B\}}( x ) Q_x \label{eq:Ufsuperposition}
\ee
we wish to show that $\mfu_{\{B\}}( x ) = \mfu_{\{B\}}(z_0)  \delta_{x,z_0}$ holds up to exponentially small corrections in system size, where $z_0\in Z(G)$. We prove this statement in two steps. We first show that $\mfu_{\{B\}}( x )$ must vanish if $x\notin Z(G)$. We then show that there can only be one term in the superposition \eqnref{eq:Ufsuperposition}. 

The first part of the proof begins by examining the commutator of $U_f$ with $V_1(y)$ as in the previous subsection. As $U_f$ is unitary, $Q_x^{\dagger}=Q_{x}^{-1}$, and the $\{B\}$ operators commute with all $Q_x$,
\begin{align}
&[[U_f:V_1(y)],F] \nonumber\\
&= \sum_{x,x'}  \mfu^{*}_{\{B\}}(x) \mfu_{\{B\}}(x') [Q_x V_1(y) Q^{-1}_{x'} V^{-1}_1(y),F]\label{eq:doublecommapp}
\end{align}
where again $F$ is chosen to be a function with support on some distant site $s$. As $U_f$ is local, any matrix elements of this commutator (with respect to  some local basis) should tend to zero exactly or exponentially fast for large $s$. Examine matrix elements $\{g'_r,\},\{g_r\}$ where $g'_1=g_1=a$. Such matrix elements  disappear on terms in the double sum \eqnref{eq:doublecommapp} unless $x=x'$.  
\begin{align}
&\langle a,\{g'_r,\}' \mid [[U_f:V_1(y)],F] \mid  a,\{g_r,\}' \rangle \nonumber\\
&= \sum_{x}  |\mfu_{\{B\}}(x)|^2 \langle a,\{g'_r,\}'\!\mid\! [Q_x V_1(y) Q^{\dagger}_{x} V^{\dagger}_1(y),F]\!\mid  \!\!a,\{g_r,\}' \rangle\nonumber\\
&= \sum_{x}  |\mfu_{\{B\}}(x)|^2  (F(g_s) - F(a x a^{-1}  y^{-1} a x^{-1}  a^{-1} y  g_s ))\nonumber \\
&\times \delta_{g'_r, a x a^{-1}   y^{-1} a x^{-1}  a^{-1}y  g_r }\label{eq:deltaconstraintapp} 
\end{align}
where again the notation $ \mid g_1, \{ g_r \}' \rangle$ isolates the group element on site $1$ from the group labels on other sites $\{g_r\}'=\{g_r:r\neq 1\}$. We now show that $|\mfu_B(x)|$ must vanish for $x=x_0$ non-abelian.  Choose
$$
F\mid \{g_r \}\rangle=\delta_{g_s,t }\mid \{g_r \}\rangle
$$
for some fixed $t\in G$, noting $F$ is clearly an operator localized to site $s$. Then the $a=1$, $g_s=t$ component of \eqnref{eq:deltaconstraintapp} becomes
$$
\sum_{x}  |\mfu_{\{B\}}(x)|^2  (1 - \delta_{x y^{-1}  x^{-1} y , 1}) \delta_{g'_r,  x y^{-1}  x^{-1} y g_r }\punc{.}
$$
Further restrict attention to the  $g'_r =   x_0 y^{-1}  x_0^{-1} y g_r $ component of \eqnref{eq:deltaconstraintapp}, obtaining
$$
\sum_{x}  |\mfu_{\{B\}}(x)|^2  (1 - \delta_{x y^{-1}  x^{-1} y , 1}) \delta_{x_0 y^{-1}  x_0^{-1} y,  x y^{-1}  x^{-1} y }\punc{.}
$$
The second delta function is non-vanishing only for a certain subset of those $x$ (which includes $x_0$) which fail to commute with $y$, hence the expression further simplifies to 
\be
\sum_{x}  |\mfu_{\{B\}}(x)|^2  \delta_{x_0 y^{-1}  x_0^{-1} y,  x y^{-1}  x^{-1} y }\punc{.}\label{eq:appx0nonab}
\ee
Using a Lieb-Robinson bound, this expression (a matrix element of a commutator) should tend to zero exponentially fast as $|s|\rightarrow \infty$. But \eqnref{eq:appx0nonab} bounds $|\mfu_{\{B\}}(x_0)|^2$ above, so that $|\mfu_{\{B\}}(x_0)|^2$ also tends to zero exponentially fast as we send $|s|\rightarrow \infty$. But \eqnref{eq:appx0nonab} is actually independent of $s$, so $|\mfu_{\{B\}}(x_0)|^2$  must be exponentially small in the system size for any non-abelian $x_0$. Hence, all $x$ appearing appreciably in the expression for $U_f$ must be in $Z(G)$ i.e.,
\be\label{eq:prefinalUfexpression}
U_f = \sum_{z\in Z(G)}  \mfu_{\{B\}}( z ) V(z)
\ee
where we used the fact $Q(z) = V(z)$ for $z \in Z(G)$ alluded to in \secref{s:nonab}.  We show that as a consequence of $U_f$ being local, $\mfu_{\{B\}}( z )$ is non-vanishing for only one $z=z_0\in Z(G)$. Recall that the operators $\mfg^{\chi}_{r,ij}$ defined in \eqnref{eq:mfgchi} obey commutation relations 
\be\label{eq:mfgchicomm2}
V(z)  \mfg^{\chi}_{r,ij}V^{-1}(z) = \frac{\chi(z)}{\chi(1)}\mfg^{\chi}_{r,ij}
\ee
because $z\in Z(G)$ acts like scalar multiplication in all irreducible representations $\chi$. Using Lieb-Robinson bounds, and the form of \eqnref{eq:prefinalUfexpression}, it follows that
\be\label{eq:Ufmfgchicomm}
U_f \mfg^{\chi}_{s, i'j'} U^{-1}_f =\mfg^{\chi}_{s, i'j'} \eta_{\chi,s}
\ee
where $\eta_{\chi,s}$ is an operator that depends only on $\chi$ and is localized around $s$. However, it follows immediately from \eqnref{eq:mfgchicomm2} that $\mfg^{\chi,\dagger}_{l,ij} \mfg^{\chi}_{s, i'j'} $ commutes exactly with $U_f$ for any sites $l,s$ however widely separated. In conjunction with \eqnref{eq:Ufmfgchicomm}, this implies
$$
\eta_{\chi,l}^{\dagger} \mfg^{\chi,\dagger}_{l, i j } \mfg^{\chi}_{s, i'j'}\eta_{\chi,s} =  \mfg^{\chi,\dagger}_{l, i j } \mfg^{\chi}_{s, i'j'}\punc{.}
$$
Using the mentioned locality properties of the operators, and $\sum_{k} \mfg^{\chi,\dagger}_{l, i k } \mfg^{\chi}_{l, k j } = \delta_{i j}$, it is readily verified that
$$
\eta_{\chi,s}^{\dagger} \eta_{\chi,l} =1\punc{.}
$$
But, as these two operators are localized very far from one another, yet inverse to one another, they must act by scalar multiplication up to exponentially small corrections in system size. Dropping the $s$ site label for now we find therefore that
$$
U_f \mfg^{\chi}_{ij} U^{-1}_f = e^{i \theta_{\chi}} \mfg^{\chi}_{ij}
$$
for all $i,j$ and irreducible representations $\chi$ where $e^{i \theta_\chi } \in \text{U}(1)$, from whence it follows

\begin{align}
 \sum_k \mfg^{\chi,\dagger}_{1k} U_f \mfg^{\chi}_{k1}  &= \sum_k   \mfg^{\chi,\dagger}_{1k}  \mfg^{\chi}_{k1}  e^{i \theta_{\chi}} U_f\nonumber\\
 &=e^{i \theta_{\chi}} U_f\nonumber \punc{.}
\end{align}
But we can evaluate the LHS of this expression using  \eqnref{eq:prefinalUfexpression} and \eqnref{eq:mfgchicomm2}, to find
\be
\sum_{z}   \mfu_{\{B\}}( z ) V(z) \frac{\chi(z) }{\chi(1)}  = \sum_{z}   \mfu_{\{B\}}( z ) V(z)  e^{i \theta_{\chi}} \punc{.}
\ee

Now, each non-vanishing term in the sum is orthogonal (use usual inner product for operators $\langle A \mid  B\rangle=\tr(A^{\dagger} B)$), so the two sums must be equal component-wise i.e.,
\be\label{eq:centereqn}
\mfu_{\{B\}}( z ) \left[\frac{\chi(z)}{\chi(1)} - e^{i \theta_{\chi}} \right]=0
\ee
for all $z\in Z(G)$ and all irreducible representations $\chi$. Suppose $\mfu_{B}$ is nonzero for some $z_0$. Then we have 
$$
\frac{\chi(z_0)}{\chi(1)} = e^{i \theta_{\chi}}
$$
for all $\chi$. Substituting this back into \eqnref{eq:centereqn} we find 
$$
\mfu_{\{B\}}( z ) \left[ \chi(z) - \chi(z_0) \right]=0
$$
for all $z\in Z(G)$ and all $\chi$. Now suppose $z_1\neq z_0$ is also in the centre. Multiplying by $\chi^*(z_1)$ and summing over $\chi$ gives (using the orthogonality relation \eqnref{eq:orthog}) 
$$
\mfu_{\{B\}}( z ) \delta_{z,z_1}=0\punc{.}
$$
Hence, $\mfu_{\{B\}}( z_1 )=0 $ for any $z_1\neq z_0$ as required. It follows therefore that 
$$
U_f = \mfu_{\{B\}}( z_0) V(z_0) \punc{,}
$$
as required.
\section{Logarithms of $V(g)$}\label{app:log}
Suppose $V$ is a unitary operator with finite order $q$. Let $\omega$ be a primitive $q^{\text{th}}$ root of unity. Here is an explicit expression for the logarithm of this operator log will take the form (for an order $N$ character)

\begin{align}
\frac{q\log(V)}{2\pi i}&= 0\delta(V=1) + 1 \delta(V=\omega^1) + \ldots +(q-1) \delta(V=\omega^{q-1})\nonumber\\
&=  \sum^{q-1}_{j=0} j \delta(V=\omega^j)\nonumber\\
&= \frac{1}{q} \sum^{q-1}_{k,j=0} j  V^{k} \omega^{-j k} \nonumber\\
&= \sum^{q-1}_{k=0}   V^{k}  c_k \label{eq:log}
\end{align}
where
$$
c_k = \frac{( (q-1) \omega^{-k (q-1)}  - q \omega^{-k q} + \omega^{k} ) }{q (\omega^{-k}-1)^2}\punc{.}
$$

\section{Order parameter correlations}\label{app:2pirot}

In this appendix we argue that FM ordered binary drives of the form \eqnref{eq:classDUf} which involve a $2 \pi$ rotation of the order parameter can be continuously deformed to drives which involve no rotation of the order parameter, without encountering an eigenstate phase transition or breaking Ising symmetry. To this end,  it is convenient to specialize to a system with an even number of site. We show that a unitary of form
\[
U(t)=e^{-it\sum_{r=1}^{N}X_{r}}\punc{,}
\]
for $0\leq t\leq\pi$ can be tuned to a constant path continuously
while maintaining Ising symmetry and fixing the endpoints $U(0)=U(\pi)=1$. This implies that binary Floquet drives of the form 
\[
U(t)=\begin{cases}
e^{-i t \sum_{r=1}^{N} X_{r}} & 0\leq t<\pi \\
e^{-iH_{1}(t-\pi)} e^{-i \pi \sum_{r=1}^{N} X_{r}} & \pi\leq t<\pi+t_{1} \punc{.}
\end{cases}
\]
where $H_1$ is potentially disordered, can be continuously tuned fixing the value of $U_{f}$, to 
\[
U(t)=e^{-iH_{1}t} \,\,\,\,\,  0\leq t\leq t_{1} \punc{.}
\]
As the system has an even number of sites, we can split
\[
U(t)=\prod_{r\text{ odd}}e^{-it(X_{r}+X_{r+1})}\punc{.}
\]
It suffices to show that for each pair of sites, we can continuously
deform 

\[
U_{r}(t)=e^{-it(X_{r}+X_{r+1})}\punc{,}
\]
to a constant unitary in an Ising symmetric manner, fixing the end
points $U_{r}(0)=U_{r}(\pi)=1$. Pick an explicit basis
for this two site system (WLOG $r=1$). 
\begin{align*}
X_{1}\otimes1_{2}= & \begin{pmatrix}1 & 0 & 0 & 0\\
0 & 1 & 0 & 0\\
0 & 0 & -1 & 0\\
0 & 0 & 0 & -1
\end{pmatrix}\\
1_{1}\otimes X_{2}= & \begin{pmatrix}1 & 0 & 0 & 0\\
0 & -1 & 0 & 0\\
0 & 0 & 1 & 0\\
0 & 0 & 0 & -1
\end{pmatrix}\\
X=X_{1}\otimes X_{2}= & \begin{pmatrix}1 & 0 & 0 & 0\\
0 & -1 & 0 & 0\\
0 & 0 & -1 & 0\\
0 & 0 & 0 & 1
\end{pmatrix}
\end{align*}

The basis is labelled by $X_1,X_2$ eigenvalues in order $11,1\bar{1},\bar{1}1,\bar{1}\bar{1}$ where $\bar{1}=-1$. At this point
it is convenient to change basis slightly ($2\leftrightarrow4$ swap)
to give 

\begin{align*}
X_{1}\otimes1_{2}= & \begin{pmatrix}1 & 0 & 0 & 0\\
0 & -1 & 0 & 0\\
0 & 0 & -1 & 0\\
0 & 0 & 0 & 1
\end{pmatrix}\\
1_{1}\otimes X_{2}= & \begin{pmatrix}1 & 0 & 0 & 0\\
0 & -1 & 0 & 0\\
0 & 0 & 1 & 0\\
0 & 0 & 0 & -1
\end{pmatrix}\\
X=X_{1}\otimes X_{2}= & \begin{pmatrix}1 & 0 & 0 & 0\\
0 & 1 & 0 & 0\\
0 & 0 & -1 & 0\\
0 & 0 & 0 & -1
\end{pmatrix}\punc{.}
\end{align*}
The basis is now ordered $11,\bar{1}\bar{1},\bar{1}1,1\bar{1}$. We wish to
find all unitaries which commute with the two site Ising symmetry
$X$. Such a unitary must take block diagonal form 
\begin{equation}
W=\begin{pmatrix}A & 0\\
0 & D
\end{pmatrix}\label{eq:Isingunitaries}
\end{equation}
with the only requirement being $A,D\in\text{U}(2)$. Now consider
the unitary $U(t)$ which in the current basis takes form
\begin{align*}
U(t)&=e^{-itX_{1}}e^{-itX_{2}}  \\
& = 
\begin{pmatrix}e^{-it} & 0 & 0 & 0\\
0 & e^{it} & 0 & 0\\
0 & 0 & e^{it} & 0\\
0 & 0 & 0 & e^{-it}
\end{pmatrix}
\times
\begin{pmatrix}e^{-it} & 0 & 0 & 0\\
0 & e^{it} & 0 & 0\\
0 & 0 & e^{-it} & 0\\
0 & 0 & 0 & e^{it}
\end{pmatrix}\\
& = 
\begin{pmatrix}e^{-2it} & 0 & 0 & 0\\
0 & e^{2it} & 0 & 0\\
0 & 0 & 1 & 0\\
0 & 0 & 0 & 1
\end{pmatrix}
\end{align*}
To see whether $U(t)$ is deformable to a constant in the space of
unitaries of form Eq.~(\ref{eq:Isingunitaries}), we need only decide
whether 

\[
A(t)=\begin{pmatrix}e^{-2it} & 0\\
0 & e^{2it}
\end{pmatrix}
\]
can be deformed to a constant, fixing its endpoints $A(0)=A(\pi)=1_{2}$,
within $\text{U}(2)$. But note that $A(t)$ lies entirely
in $\text{SU}(2)\subset\text{U}(2)$ because $\det A(t)=1$. As $\text{SU}(2)$
is a simply connected space i.e, $\pi_{1}(\text{SU}(2))=\{1\}$, it
must be the case that $A(t)$ can be continuously deformed to a constant
while fixing its endpoints. In other words, the closed loop defined
by $A(t)$ lies entirely in simply connected space $\text{SU}(2)$,
and can thus be deformed to a point.  We can WLOG reparameterize this unitary as

\[
A(t)=\begin{pmatrix}e^{-it} & 0\\
0 & e^{it}
\end{pmatrix}
\]
$0\leq t\leq2\pi$. To deform this to a constant path, use an interpolating family of unitaries

\[
A(t;\lambda)=\begin{pmatrix}(e^{-it}-1)c_{\lambda}^{2}+1 & c_{\lambda}s_{\lambda}(1-e^{-it})\\
-c_{\lambda}s_{\lambda}(1-e^{it}) & (e^{it}-1)c_{\lambda}^{2}+1
\end{pmatrix}\punc{,}
\]
where $\lambda \in [0,1]$, $c_\lambda = \cos (\pi \lambda/2)$, and $s_\lambda = \sin (\pi \lambda/2)$. Then $A(t,0)=A(t)$ while $A(t,1)=1$ as required.
\end{appendix}

%\bibliography{global}
\end{document}